\documentclass[10pt,a4paper]{article}

\usepackage{authblk}
\usepackage{url}
\usepackage{amsmath,amssymb,amsthm}
\usepackage{graphicx}
\usepackage{caption}
\usepackage{flushend}
\usepackage{cite}
\usepackage{nameref,hyperref}
\usepackage{xcolor}
\usepackage{pdfpages}
\usepackage[normalem]{ulem}
\usepackage[margin=1.3in]{geometry}

\newcommand{\xx}{\mathbf{x}}

\newcommand{\pmi}{{\sf pmi}}

\newcommand{\edm}{{\sf Edmonds}}
\newcommand{\prim}{\textsf{Prim}}
\newcommand{\gbw}{\textsf{Gabow}}
\newcommand{\chl}{\textsf{Chow-Liu}}
\newcommand{\epos}{\epsilon_+}
\newcommand{\eneg}{\epsilon_-}

\newcommand{\SCITE}{\textsf{SCITE}}

\newcommand{\OncoNEM}{\textsf{OncoNEM}}

\newcommand{\CHOWLIU}{\textsf{CHOW-LIU}}
\newcommand{\PRIM}{\textsf{PRIM}}
\newcommand{\EDMONDS}{\textsf{EDMONDS}}
\newcommand{\GABOW}{\textsf{GABOW}}

\newcommand{\TRONCO}{\textsf{TRONCO}}

\newcommand{\TRAIT}{\textsf{TRaIT}}

\newcommand{\rev}[1]{\textcolor{black}{#1}}
\newcommand{\revfinal}[1]{\textcolor{black}{#1}}

\newcommand{\GPF}{$G_{\mathsf{PF}}$}
\newcommand{\GI}{$G_{\mathsf{NL}}$}
\newcommand{\GMO}{$G_{\mathsf{MO}}$}
\renewcommand{\sout}[1]{}

\begin{document}

\title{\text{L}earning mutational graphs of individual tumour evolution from single-cell and multi-region sequencing data}
 
\author[1]{Daniele Ramazzotti}
\author[2,3]{Alex Graudenzi\footnote{To whom correspondence should be addressed.}}
\author[2]{Luca De Sano}
\author[2,4]{Marco Antoniotti}
\author[5]{Giulio Caravagna}
\affil[1]{Department of Pathology, Stanford University, Stanford, CA 94305, USA}
\affil[2]{Dipartimento di Informatica, Sistemistica e Comunicazione, Universit\`{a} degli Studi di Milano-Bicocca, 20126 Milan, Italy}
\affil[3]{Institute of Molecular Bioimaging and Physiology of the Italian National Research Council (IBFM-CNR), Viale F.lli Cervi 93, 20090 Segrate, Milan, Italy}
\affil[4]{Milan Center for Neuroscience, Universit\`{a} degli Studi di Milano-Bicocca, Via Pergolesi 33, 20052 Monza, Italy}
\affil[5]{Centre for Evolution and Cancer, The Institute of Cancer Research, 15 Cotswold Road, SM2 5NG London, United Kingdom}

\date{}

\maketitle

\begin{abstract}
\hspace{-0.5cm}\textbf{Background.} A large number of algorithms is being developed to reconstruct evolutionary models of individual tumours from genome sequencing data. Most methods can analyze multiple samples collected either through {bulk multi-region} sequencing experiments or the sequencing of individual cancer cells. \revfinal{However, rarely the same method can support both data types.} \\
\textbf{Results.} We introduce \TRAIT{}, a computational framework to infer mutational graphs that model the accumulation of multiple types of somatic alterations driving tumour evolution.  Compared to other tools, \TRAIT{} supports multi-region and single-cell sequencing data within the same statistical framework, \revfinal{and delivers expressive models that capture many complex evolutionary phenomena.} \TRAIT{} improves accuracy, robustness to data-specific errors and computational complexity compared to  competing methods. \\
\textbf{Conclusions.} We show that the application of \TRAIT{} to single-cell and multi-region cancer datasets can produce accurate and reliable models of single-tumour evolution, quantify the extent of intra-tumour heterogeneity and generate new testable experimental hypotheses. 
\end{abstract}

\section*{Background}
\label{sect:intorduction}

\textcolor{black}{Sequencing data from multiple samples of single tumours can be used to investigate {Intra-Tumor Heterogeneity} (ITH) in light of evolution \cite{beerenwinkel2015cancer,schwartz2017evolution,RepeatedCaravagna2018}. Motivated by this observation, several new methods  have been developed to infer the ``evolutionary history'' of a  tumour from sequencing data.  According to  Davis and Navin, there are three orthogonal ways  to depict such history \cite{davis2016computing}: $(i)$ with a phylogenetic tree that displays input samples as leaves  \cite{stamatakis2014raxml}, $(ii)$ with  a  clonal tree of  parental relations between putative cancer  clones \cite{yuan2015bitphylogeny,ross2016onconem,roth2016clonal,zafar2017sifit}, and $(iii)$ with the order of mutations that accumulated during cancer growth \cite{kim2014using,jahn2016tree,salehi2017ddclone}. Ideally, the order of accumulating mutations should match the clonal lineage tree in order to reconcile  these inferences.}  \textcolor{black}{Consistently with earlier works of us \cite{loohuis2014:_caprese_infer,ramazzotti2015capri,Caravagna2016E4025,ramazzotti2017model,ramazzotti2016parallel,ramazzotti2017modeling}, we here approach the third problem (``mutational ordering'')  from 
 two  types of data: {multi-region bulk} and {single-cell} sequencing.} 

Bulk sequencing of multiple spatially-separated tumour biopsies  returns a noisy mixture of admixed lineages  
\cite{gerlinger2012intratumor,de2014spatial,zhang2014intratumor,yates2015subclonal,jamal2017tracking}. We can analyse these data by first retrieving clonal prevalences in bulk samples (subclonal deconvolution),  and then by computing their
evolutionary relations 
\cite{oesper2013inferring,strino2013trap,roth2014pyclone,jiao2014inferring,zare2014inferring,deshwar2015phylowgs,el2016inferring,hu2016inferring}.  Subclonal deconvolution is usually computationally challenging, and can be avoided if we can read genotypes of individual cells via single-cell sequencing (SCS). Despite this theoretical advantage, however, current technical challenges in cell isolation and genome amplification are major bottlenecks to scale  SCS to   whole-exome or whole-genome assays, and the available targeted data  harbours  high levels of  {allelic dropouts}, {missing data} and {doublets}\cite{navin2011tumour,gawad2014dissecting,wang2014clonal,navin2015first}. Thus,  the direct application of standard phylogenetic methods to SCS data is not straightforward, despite being theoretically viable \cite{schwarz2014phylogenetic}. 

\revfinal{Notice that a common feature of most  methods for cancer evolution reconstruction is the employment of the {Infinite Sites Assumption} (ISA), together with the assumption of no back mutation \cite{oesper2013inferring,strino2013trap,roth2014pyclone,jiao2014inferring,zare2014inferring,deshwar2015phylowgs,el2016inferring,hu2016inferring,navin2011tumour,gawad2014dissecting,wang2014clonal,navin2015first}, even though recent attempts (e.g., \cite{zafar2017sifit}) have been proposed to relax such assumption in order to model relevant phenomena, such as convergent evolutionary trajectories \cite{navin2016genotyping}.} 

\revfinal{\sout{Our contribution is the definition of a new statistical framework (and a set of different algorithms) which takes into account the above considerations without exposing the user to  complex setups and computational requirements. }} 

\revfinal{In this expanding field, we here introduce \TRAIT{} (\textsf{{T}emporal o{R}der of {I}ndividual {T}umors} -- Figures \ref{fig:approach} and \ref{fig:framework}), a new framework for the inference of models of single-tumour evolution, which can analyse, separately, multi-region bulk and single-cell sequencing data, and which allows to capture many complex evolutionary phenomena underlying cancer development.} 

Compared to other approaches that might scale poorly for increasing sample sizes, our methods show excellent computational performance and scalability, rendering them suitable to anticipate the large amount of genomic data that is becoming increasingly available. 

\section*{Results}
\label{sect:results}

\begin{figure}[p]
 \centerline{\includegraphics[width=1\textwidth]{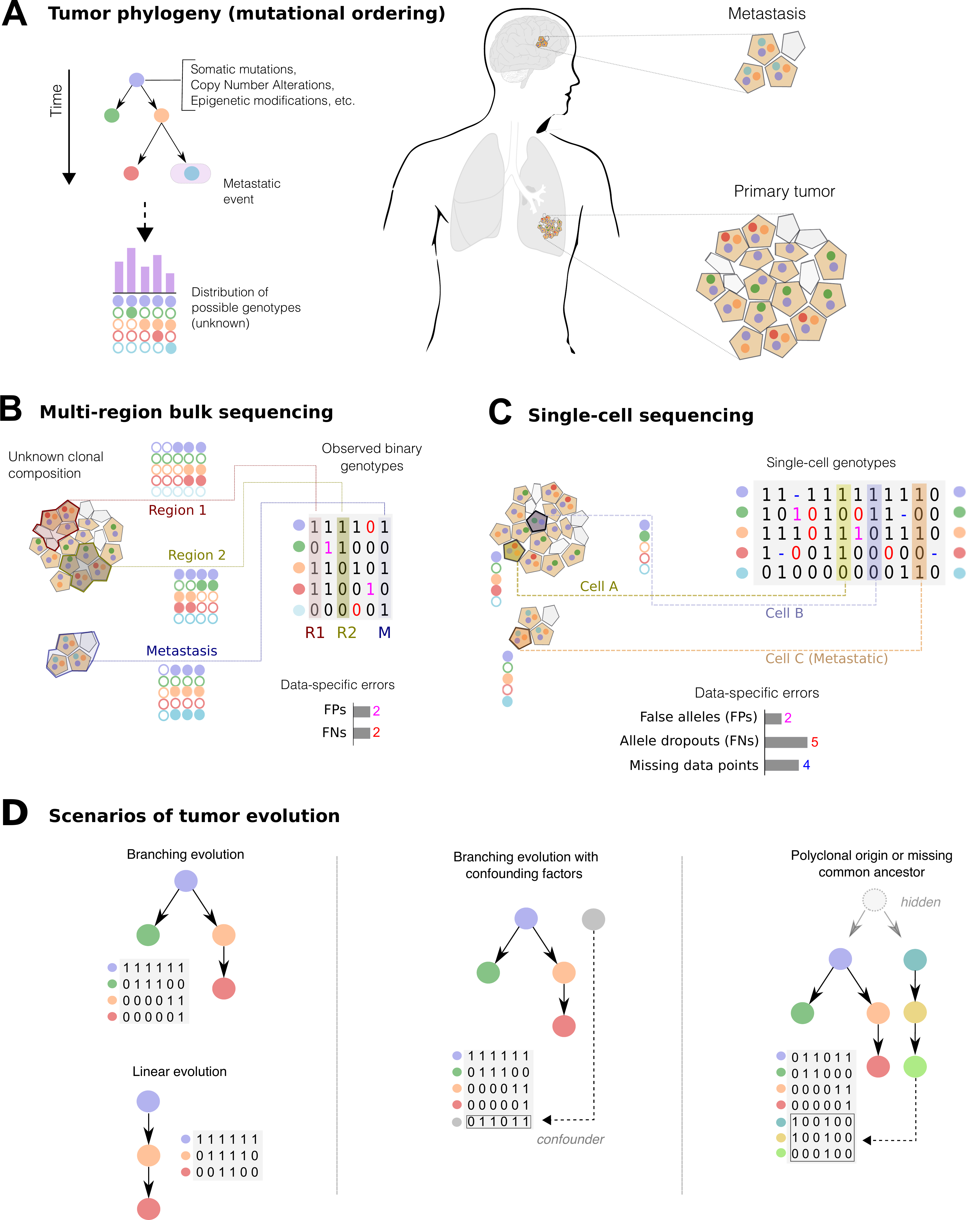}}
\caption[Approach]{\textbf{A.} A tumour phylogeny describes the order
  of accumulation of somatic mutations, CNAs, epigenetic
  modifications, etc. in a single tumour. The model generates
  a set of possible genotypes, which are observed with an
  unknown  spatial and density distribution in a tumour (primary and 
  metastases).
  \textbf{B.} Multi-region bulk sequencing returns a mixed signal from
  different  tumour subpopulations, with potential contamination of
  non-tumour cells (not shown) and symmetric rates
  of false positives and negatives in the calling. Thus, a sample will harbour 
  lesions from different tumour lineages, creating
  spurious correlations in the data.   %
  \textbf{C.} If we sequence genomes of single cells we can, in
  principle, have a precise signal from each subpopulation. However,
  the  inference with these data is made harder by high levels of
  asymmetric noise, errors in the calling and missing data. 
  \textbf{D.} Different scenarios of tumour evolution 
  can be investigated via \TRAIT{}. $(i)$ Branching evolution (which 
  includes linear evolution), $(ii)$
  Branching evolution with confounding factors annotated in the
  data, $(iii)$ Models with multiple progressions due to
  polyclonal tumour origination, or to the presence tumour initiating event
  missing from input data \revfinal{\sout{and $(iv)$ generalised convergent evolution (ISA violation)}.}}
\label{fig:approach}
\end{figure}

\begin{figure*}[t]
\centerline{\includegraphics[width=1\textwidth]{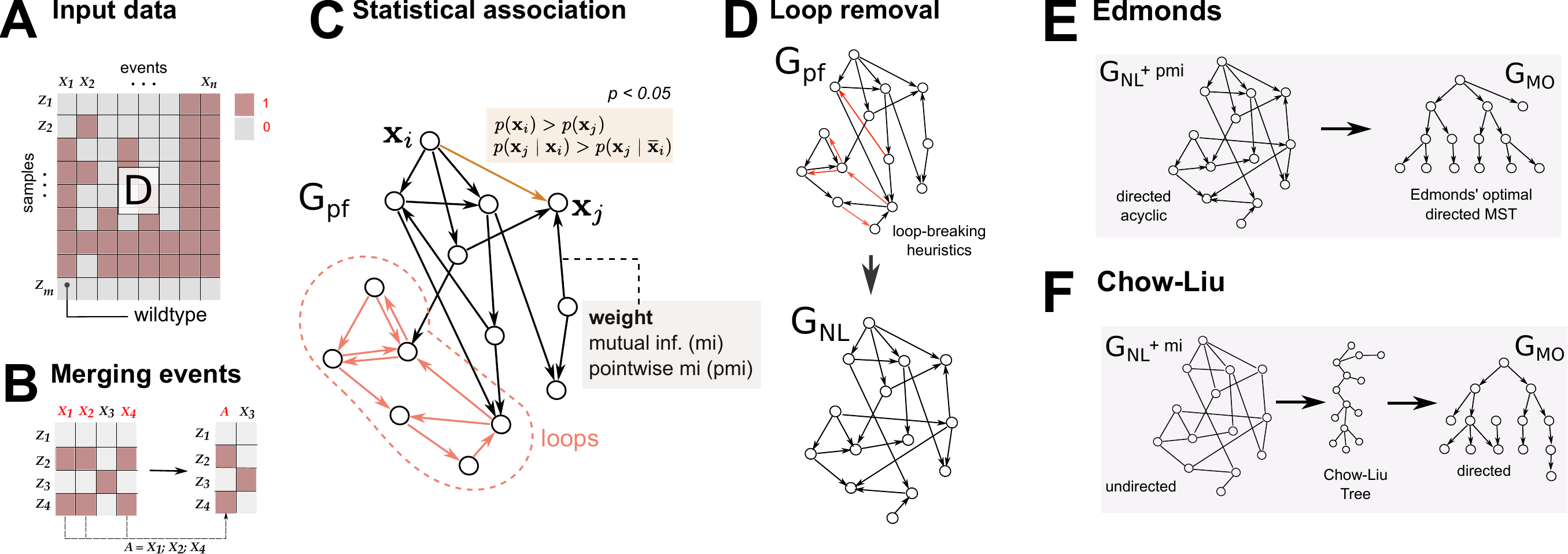}}
  \caption[A computational framework for individual-level inference]{
    \textbf{A}. \TRAIT{} processes a binary matrix $D$ that stores
    the presence or absence of a variable in a sample (e.g., a mutation, a
    CNA, or a persistent epigenetic states). 
    \textbf{B.} \TRAIT{} merges the events occurring in the same samples 
    ($\xx_1$, $\xx_2$ and $\xx_4$, merged to A), as the statistical signal for their
    temporal ordering is  undistinguishable.
    The final model  include such aggregate events.
    \textbf{C}. We estimate via bootstrap the \emph{prima facie}
    ordering relation that satisfies Suppes' conditions
    (\ref{eq:sup}) for statistical association. This induces a graph \GPF{}
    over variables $\xx_i$, which is weighted by
    information-theoretic measures for variables' association such as
    mutual information or pointwise mutual information.
    \textbf{D.} \TRAIT{} employs heuristic strategies  to remove loops
    from  \GPF{} and produce a new  graph \GI{} \cite{ramazzotti2015capri}.
    \textbf{E.} \edm{}'s algorithm can be used to reconstruct the
    optimal minimum spanning tree \GMO{} that minimises the
    weights in \GI{}; here we use point-wise mutual information (\pmi{}).
    \textbf{F.} \chl{} is a Bayesian  mode-selection strategy
    that  computes an undirected tree as a model of a joint distribution on
    the annotated variable. Then, we
    provide edge direction (temporal priority), with Suppes'
    condition (\ref{eq:sup}) on marginal probabilities. Therefore,
    confluences are possible in the output model \GMO{} in certain
    conditions \revfinal{\sout{, and this allows for violations of the ISA}.}}
  \label{fig:framework}
\end{figure*}

\textcolor{black}{\TRAIT{} is a computational framework that combines Suppes' {probabilistic causation} \cite{suppes1970probabilistic}  with  information theory to  infer the temporal ordering of mutations that accumulate during tumour growth, as an extension  of our previous work \cite{loohuis2014:_caprese_infer,ramazzotti2015capri,Caravagna2016E4025,ramazzotti2017model,ramazzotti2016parallel,ramazzotti2017modeling}.  The framework comprises $4$ algorithms (\EDMONDS{}, \GABOW{}, \CHOWLIU{} and \PRIM{}) designed to model different  types of progressions ({expressivity}) and integrate various types of data, still maintaining a low burden of {computational   complexity} (Figures \ref{fig:approach} and \ref{fig:framework} -- see Methods for the algorithmic details). 
} 

In \TRAIT{}  we estimate  the {statistical association} between a set of  genomic
{events} (i.e., mutations, copy number, etc.) annotated in sequencing data by combining optimal graph-based  algorithms with bootstrap, hypothesis testing and information theory (Figure \ref{fig:framework}).   \revfinal{\sout{By using  optimal  algorithms, we can extend canonical mutational  trees to mutational directed  acyclic graphs (DAGs), a more general class of progression models  where {confluences} can suggest violations of the ISA. Furthermore, the  framework also supports models with disconnected components (i.e., multiple roots).}} 
\revfinal{\TRAIT{} can reconstruct trees and forests -- in general, {\it mutational graphs} --  which in specific cases can include confluences, to account for the uncertainty on the precedence relation among certain events.  
Forest models (i.e., disconnected trees), in particular,} can  stem for   possible  {polyclonal tumour initiation} (i.e., tumours with  multiple cells of  origin  \cite{parsons2008many}), or  the presence of  tumour-triggering events that are not annotated in the input data (e.g., epigenetic events) (Figure \ref{fig:approach}D). 

Inputs data in \TRAIT{} is represent as binary vectors, which  is the standard representation for SCS sequencing and is hereby used to define a unique   framework for both  multi-region bulk and SCS data (Figure \ref{fig:approach}A--C).  For a set of cells or regions sequenced, the input reports the presence/absence of $n$  genomic events, for which \TRAIT{}  will layout a temporal ordering. A binary representation allows to include  several {types} of somatic lesions in the analysis, such as somatic mutations (e.g., single-nucleotide, indels, etc.), copy number alterations, epigenetic states (e.g., methylations, chromatin modifications), etc. \revfinal{(see the Conclusions for a discussion on the issue of data resolution).}

\subsection*{Performance evaluation with synthetic simulations}
\label{sect:simulations}

We assessed the performance of \TRAIT{} with both SCS and multi-region data simulated from different types of generative models. 

\revfinal{\sout{with $5\leq
n \leq 20$ nodes, accounting for the tumour evolution scenarios displayed in Figure \ref{fig:approach}D (see Methods for further details).
Each node  represents one (or more) somatic lesions detected in the same samples, and for which we wish to retrieve a temporal ordering. We simulated samplings  of   $10 \leq m \leq 100$ cells in SCS assays, and  $5 \leq m \leq 50$ regions (accounting for bulk sampling bias) in   multi-region ones.}} 

\paragraph{\revfinal{Synthetic data generation.}} Synthetic single-cell datasets were sampled from a large number of randomly generated topologies (trees or forests) to reflect \TRAIT{}'s generative model. For each generative topology, binary datasets were generated starting from the root, with a recursive procedure which we describe for the simpler case of a tree: $(i)$ for the root node $x$, the corresponding variable is assigned $1$ with a randomly sampled probability $p(x=1) = r$, with $r\sim U[0,1]$; $(ii)$ given a branching node $y$ with children $y_1, y_2, \dots, y_n$, we sample values for the $n$ variables $y_1, y_2, \dots, y_n$ so that at most one randomly selected child contains $1$, and the others are all $0$. The recursion proceeds from the root to the leaves, and stops whenever a $0$ is sampled or a leaf is reached. \revfinal{Note that we are simulating exclusive branching lineages, as one expects from the accumulation of mutations in single cells under the ISA.}

As bulk samples usually include intermixed tumour sub-populations, we simulated bulk datasets by pooling single-cell genotypes generated as described above, and setting simulated variables (i.e., mutations) to $1$ (= present) in each bulk sample if they appear in the sampled single-cell genotypes more than a certain threshold. More details on these procedures are reported in \revfinal{Section 2 of the Supplementary Material.} 

\begin{figure}[p]
 \centerline{\includegraphics[width=1\textwidth]{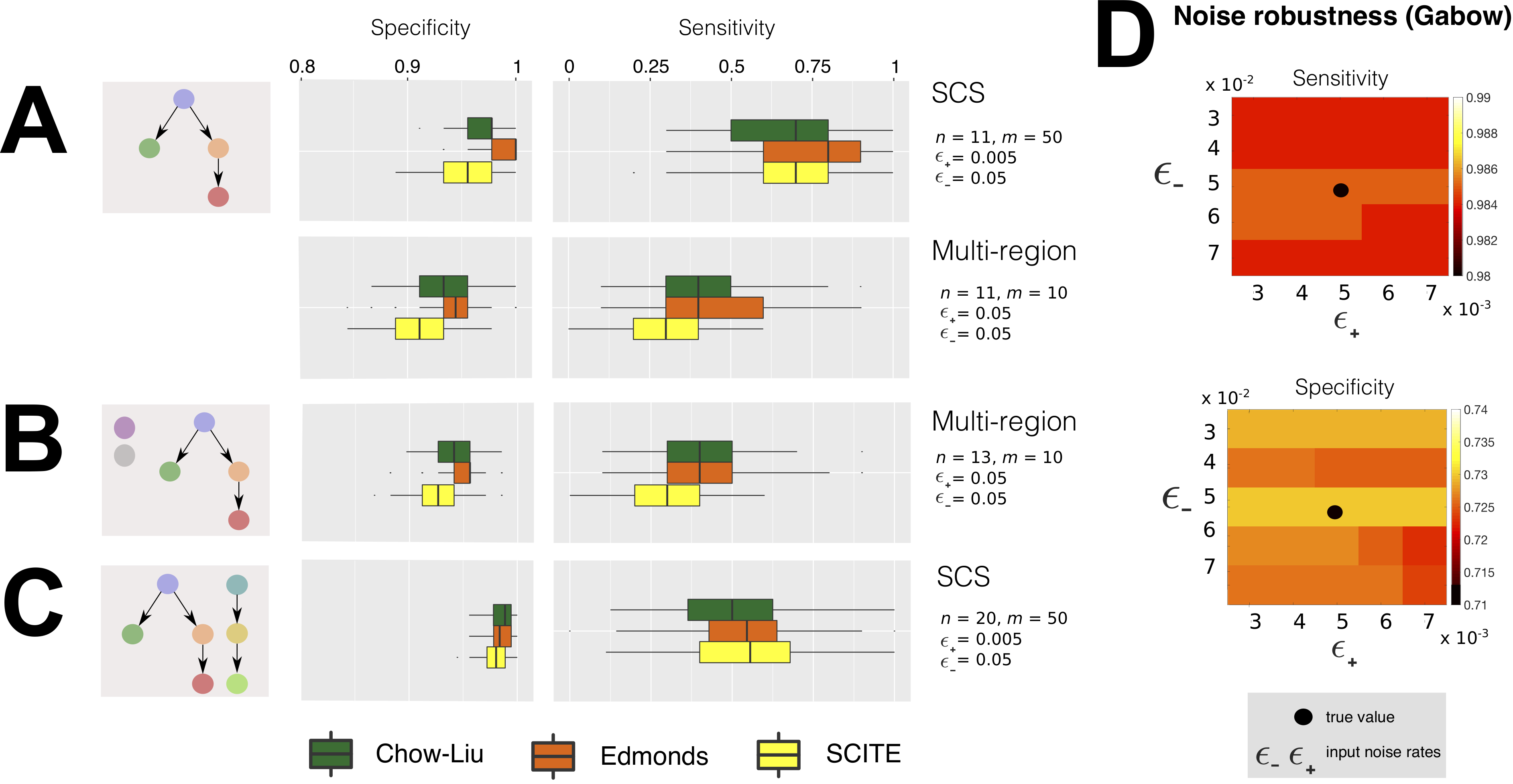}}
  \caption[Performance]{We estimate from  simulations the rate of
    detection of true positives ({sensitivity}) and negatives
    ({specificity}), visualised  as \emph{box-plots} from $100$
    independent points each. We compare \TRAIT{}'s algorithms \edm{} and
    \chl{}  with \SCITE{}, the state-of-the-art for mutational trees
    inference in a setting of mild noise in the data, and canonical sample size. 
    In SCS data noise is
    $\epsilon _{+} = 5 \times 10^{-3}; \epsilon _{-} = 5 \times 10^{-2}$,
    in multi-region $\epsilon _{-} = 5 \times 10^{-2}$.  Extensive
    results for different models, data type, noise and sample size are
    in Supplementary Figures S3--S16.
    \textbf{A.} Here we use a generative model from
    \cite{yuan2015bitphylogeny} (Supplementary Figure S7-B). (left)
    SCS datasets with $m=50$ single cells, for a tumour with $n=11$
    mutations. (right) Multi-region datasets with $m=10$ spatially
    separated regions, for a tumour with  $n=11$ mutations.
    \textbf{B.} We augment the setting in {A-right} with $2$ random
    variables (with random marginal probabilty) to model confounding
    factors, and generated SCS data.
    \textbf{C.} We generated  multi-region data from a  tumour with
    $n=21$ mutations, and a random number of  $2$ or $3$   distinct
    cells of origin to model polyclonal tumour origination.
    \textbf{D.}  Spectrum  of average sensitivity and specificity for
    \gbw{} algorithm included in \TRAIT{} (see SM) estimated from
    $100$ independent SCS datasets sampled from the generative model
    in Supplementary Figure S7-B ($m=75$, $n=11$). The true noise
    rates are $\epsilon _{+} = 5 \times 10^{-3}; \epsilon _{-} = 5
    \times 10^{-2}$; we scan input
    $\epos$ and $\eneg$ in the   ranges:  $\epos = (3, 4, 5, 6, 7)
    \times 10^{-3}$ and $3 \times 10^{-2} \leq \eneg =\leq 7 \times 10^{-2}$.}
\label{fig:performance}
\end{figure}

Consistently with previous studies, we also introduced noise in the true genotypes via inflated  false positives and false negatives, which are assumed to have highly asymmetric rates for SCS data. For SCS data we also included  missing data in a proportion of the simulated variables \cite{jahn2016tree}. 
\revfinal{Notice that \TRAIT{} can be provided with input noise rates, prior to the inference: therefore, in each reconstruction experiment we provided the algorithm with the noise rates used to generate the datasets, even though mild variations in such input values appear  not to affect the inference accuracy -- as shown in the noise robustness test presented below and in Figure \ref{fig:performance}\textsf{D}}. 

With a total of $\sim$140.000 distinct simulations,  we could reliably estimate   the ability  to infer true edges ({sensitivity}) and discriminate false ones ({specificity}); \revfinal{further details on parameter settings are available in Section 6 of the Supplementary Material.} 

\revfinal{In particular, we compared \TRAIT{}'s algorithms to  \SCITE{}, the state-of-the-art to infer {mutational trees} from SCS data \cite{jahn2016tree}. We could not include \OncoNEM{}\cite{ross2016onconem} -- the benchmark tool for clonal deconvolution -- in the comparison, as its computational performance did not scale well with our large number of tests.} 

\revfinal{In the Main Text we show  results for the \edm{} and \chl{} algorithms, included in \TRAIT{}, and \SCITE{}, in a selected number of relevant experimental scenarios.  To improve  readability of the manuscript, we leave to the Supplementary Material  a  comprehensive presentation of the results for \gbw{}, \prim{} and other approaches \cite{loohuis2014:_caprese_infer,ramazzotti2015capri}.}

\revfinal{\sout{In the branching evolution scenario (one single generative tree -- Figure \ref{fig:performance}A), all the techniques achieve high sensitivity and specificity  with SCS, and  lower scores with multi-region data from the same topology;  \edm{} achieved in general the best results with SCS data (medians $\sim0.8$ and $\sim1$).}} 

\paragraph{\revfinal{Results from scenario (i), branching evolution.}} \revfinal{To simulate  branching evolution  \cite{gerlinger2012intratumor}, we generated a large number of independent datasets from single-rooted tree structures. In particular, we employed three control polyclonal topologies taken from \cite{yuan2015bitphylogeny} (Supplementary Figure 7) and $100$ randomly generated topologies, with a variable number of nodes (i.e., alterations) in the range $n \in [5;20]$.  
Such generative models were first used to sample datasets with different number of sequenced cells ($m=10,50,100$). In addition to the noise-free setting, we perturbed data by introducing plausible and highly asymmetric noise rates (i.e., $\epos=\eneg=0$ ({\it noise-free}); $\epos = 0.005$, $\eneg = 0.05$; $\epos = 0.02$, $\eneg =  0.2$.). 
The same generative topologies were then used to sample multi-region datasets with different number of regions ($m=5,10,20$), and symmetric noise rates ($\epos = \eneg = 0, 0.05, 0.2$). } 

\revfinal{In Figure \ref{fig:performance}A we show two selected experimental settings, which are characteristic of the general trends observed on all tests. In particular, one can notice that all the techniques achieve high sensitivity and specificity with SCS data, and significantly lower scores with multi-region data from the same topology;  \edm{} displays in general the best results with SCS data (medians $\sim0.8$ and $\sim1$).}

\revfinal{From the results in all simulation settings (Supplementary Figures 8 and 9 for the SCS case; Supplementary Figures 13 and 14 for the multi-region case), we observe that the overall performance significantly improves for lower noise levels and larger datasets across for all the algorithms, a general result that is confirmed in the other experimental scenarios.}
\revfinal{In particular, with SCS data, \edm{} and \SCITE{} display similar sensitivity, even though the latter presents (on average) lower specificity, which might point to a mild-tendency to {overfit}.
Results on multi-region data display similar trends, with \edm{} showing the overall best performance and  \SCITE{} showing slightly lower performance, especially with small datasets and/or low noise levels. }

\revfinal{We also specify that, as \TRAIT{}'s algorithms share the same constraints in the search space and several algorithmic properties, the reduced variance observed across settings is expected.} 

\paragraph{\revfinal{Results from  scenario (ii), confounding factors.}} \revfinal{To investigate the impact of possible confounding factors on inference accuracy, we introduced in the datasets from scenario $(i)$ a number of random binary variables totally unrelated to the progression. More in detail, we inserted around $n \times 10 \%$ additional random columns in all datasets with $n$ input variables; each additional column is a repeated sampling of a biased coin, with bias uniformly sampled among the marginals of all events. }

\revfinal{The performance of \TRAIT{} and \SCITE{} in a selected setting for the multi-region case is shown in  Figure \ref{fig:performance}B}. Surprisingly, the introduction of confounding factors   does not  impact the performance significantly. In fact, despite two extra variables annotated in the data that are unrelated to the progression, 
most algorithms still discriminate the true generative model.
\revfinal{Similar results are achieved in the SCS case (Supplementary Figure 10).}

\paragraph{\revfinal{Results from  scenario (iii), forest models.}} \revfinal{Forest topologies can be employed as generative models of tumours initiated by multiple cells, or of tumours whose initiation is triggered by events that are not annotated in the input data. In this test we randomly generated forests with a variable number of distinct disconnected trees, thus assuming that no mutations are shared across the trees. In detail,  we generated $100$ random forest topologies, with  $n = 20$ nodes and $q < 5$ distinct roots (i.e., disconnected trees), both in the SCS and the multi-region case. }

\revfinal{The performance of the tested algorithms in a selected experimental scenario with SCS is shown in  Figure \ref{fig:performance}C. All algorithms display a clear decrease in sensitivity, with respect to the single-rooted case with similar values of noise and sample size. }
\revfinal{In the SCS case the performance remarkably increases with larger datasets (median values $\sim 0.75$ with $m = 100$ samples in the noise-free case;  Supplementary Figure 11).  \edm{} shows the best tradeoff between sensitivity and specificity, whereas \SCITE{} confirms a mild tendency to overfit for small datasets, yet being very robust against noise. Results from multi-region analysis show an overall decrease in performance (Supplementary Figure 16).}

\revfinal{\sout{However, when the generative model contains a mixture of  independent
trajectories (i.e., polyclonal origin), sensitivity decreases similarly across all methods (Figure \ref{fig:performance}C). }}

\paragraph{\revfinal{Robustness to variations in noise input values.}} Similarly to other tools, e.g., \cite{ross2016onconem,jahn2016tree}, our algorithms can receive rates of false positives and negatives
in the data ($\epos$ and $\eneg$) as input. Thus, we  analyzed the effect of 
miscalled  rates on the overall performance. \revfinal{\sout{and found  stable  values  across all
algorithms}}
\revfinal{More in detail, we analyzed the variation of the performance of \gbw{} and \SCITE{}, on a  dataset generated  from a generative tree  with intermediate complexity ({\it ``Medium"} topology in Supplementary Figure $7$), with $n=11$ nodes and $m=75$ samples,
  $\epos = 5 \times 10^-3$ and  $\eneg = 5 \times 10^-2$. We scanned   
  $25$ possible combinations of input  $\epos$ and $\eneg$ in the
  following ranges: $\epos = (3, 4, 5, 6, 7) \times 10^{-3}$ and
  $\eneg = (3, 4, 5, 6, 7) \times 10^{-2}$.}
\revfinal{Results in Figure \ref{fig:performance}\textsf{D} and Supplementary Tables 4 and 5 show no significant variations of the performance with different combinations of input values for $\epos$ and $\eneg$, for both  algorithms. This  proves that the accuracy of the inference is robust to variations in the noise input values, as long as they are reasonably close to the real value. }
This evidence also supports our algorithmic design choice which  avoids   sophisticate noise-learning strategies in \TRAIT{},
a further reason that speeds up computations.

\paragraph{\revfinal{Missing data.}} \revfinal{Significant rates of missing data are still quite common in SCS datasets, mainly due to amplification biases during library preparation. We evaluated the impact of missing data by using $20$ benchmark single-cell datasets which were generated from a  tree with $n=11$ nodes (Supplementary Figure 7). For every dataset we simulated the calling of mutations from  $m=75$ single sequenced cells, and in half of the cases (i.e., $10$ datasets) we also  imputed extra  error rates in the data to model sequencing errors. In particular, we introduced false positives and false negative calls with rates  $\epos = 0.005$ and $\eneg=0.05$. On top of this, for each of the $20$ datasets we generated $5$ configurations of missing data (uniformly distributed), using as measure the percentage $r$ of missing data over the total number of observations. A total  of $100$ distinct datasets have been obtained using  $r=0, 0.1, 0.2, 0.3, 0.4$ (i.e., up to 40\% missing data). As \SCITE{} can explicitly learn parameters from missing data,  we run  the tool with no further parameters. Instead, for \TRAIT{}'s algorithms, we performed the following procedure: for each dataset $D$ with missing data, we imputed the missing entries via a standard Expectation-Maximization (EM) algorithm, repeating the procedure  to generate $100$ complete  datasets ($D_1, \dots, D_{100}$). To asses the performance of each algorithm, we computed the fit to all the $100$ datasets, and selected the solution that maximised the likelihood  of the model.}

\begin{figure}[p]
 \centerline{\includegraphics[width=1\textwidth]{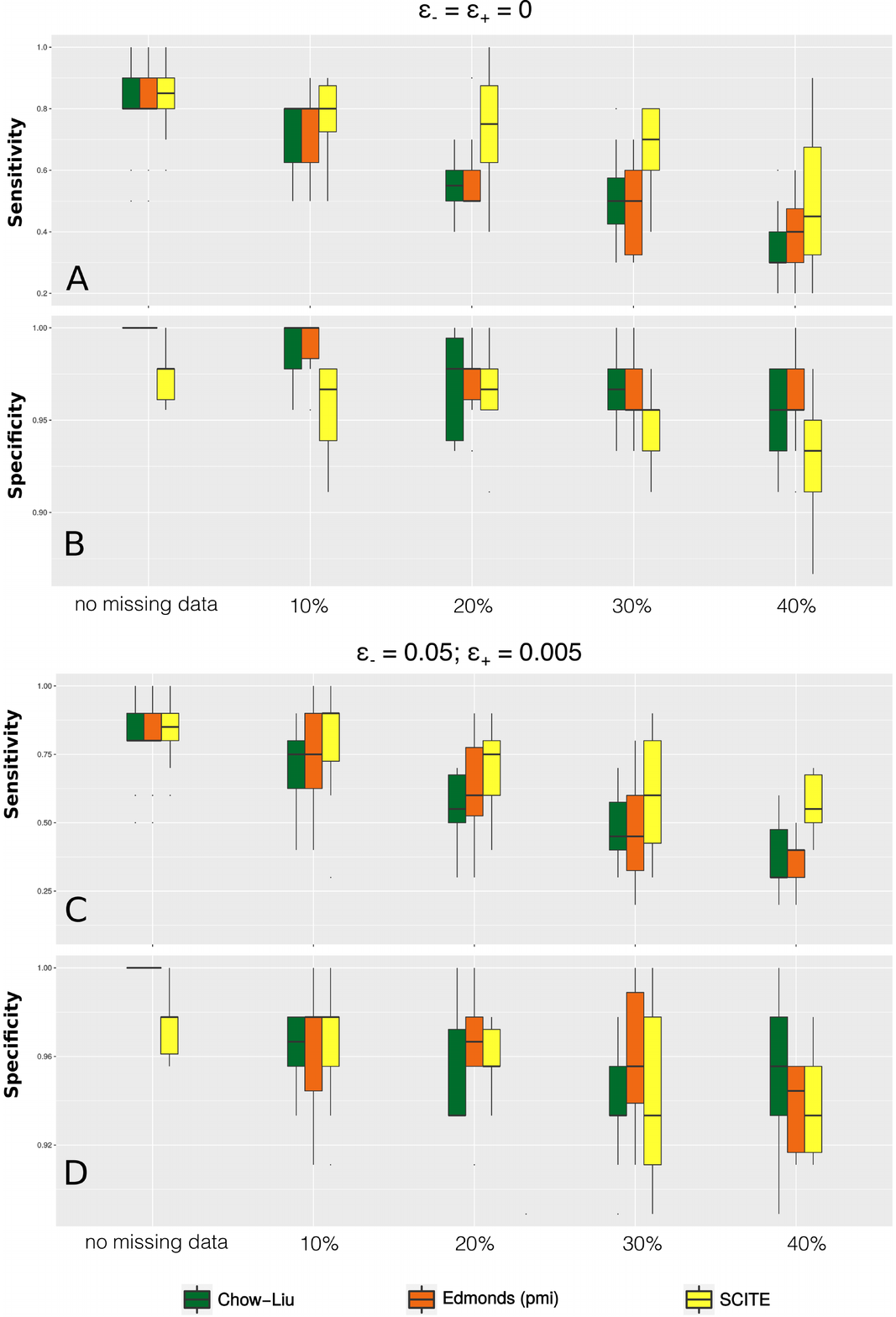}}
\caption[Performance on SCS with missing data)]{\revfinal{Sensitivity and specificity for different percentages $r$ of missing entries, namely, $r=(0, 0.1, 0.2, 0.3, 0.4)$ as a function of the number of variables in the data, and different levels of noise: $(i)$ $\epos=\eneg=0$ and $(ii)$ $\epos=0.005$, $\eneg=0.05$. The original dataset is generated from a tree with $n=11$ nodes and $m=75$ samples (Supplementary Figure 7).}}
\label{fig:performance_md}
\end{figure}

\revfinal{We present in Figure \ref{fig:performance_md} the results of this analysis for \edm{} and \chl{} algorithms included in \TRAIT{}, and for \SCITE{};  results for  \gbw{} and \prim{} algorithms are presented in  Supplementary Figure $12$.  In general, missing data profoundly affect the performance of all methods. \SCITE{} shows overall more robust sensitivity, in spite of slightly worse specificity. The performance  is always significantly improved when data do not harbour noise and, in general, is reasonably robust up to $30\%$ missing data.}

\paragraph{\revfinal{Computational time.}} \revfinal{One of the major computational advantages of \TRAIT{} is its scalability, which will be essential in anticipation of the increasingly larger SCS datasets expected in the near future. In this respect, we have observed  across all tests a $3\times$ speedup of \TRAIT{}'s algorithms on standard CPUs with respect to \SCITE{}, \revfinal{and  a $40\times$ speedup with respect to \OncoNEM{}} (Supplementary Table 6).}

\revfinal{\sout{Overall conclusions can be drawn from all the tests. As
expected, performances improve with lower noise and larger datasets.
In particular, with SCS data \edm{} and \SCITE{} display very similar
sensitivity, even though the latter  presents (on average) lower
specificity. 
Results on multi-region data display similar trends, with
\edm{} showing the overall best performance and  \SCITE{} showing slightly lower performance, 
especially with small datasets and/or low noise levels. }}
\\

\subsection*{Analysis of patient-derived multi-region data for a MSI-high colorectal cancer}

\begin{figure}[t]
 \centerline{\includegraphics[width=1\textwidth]{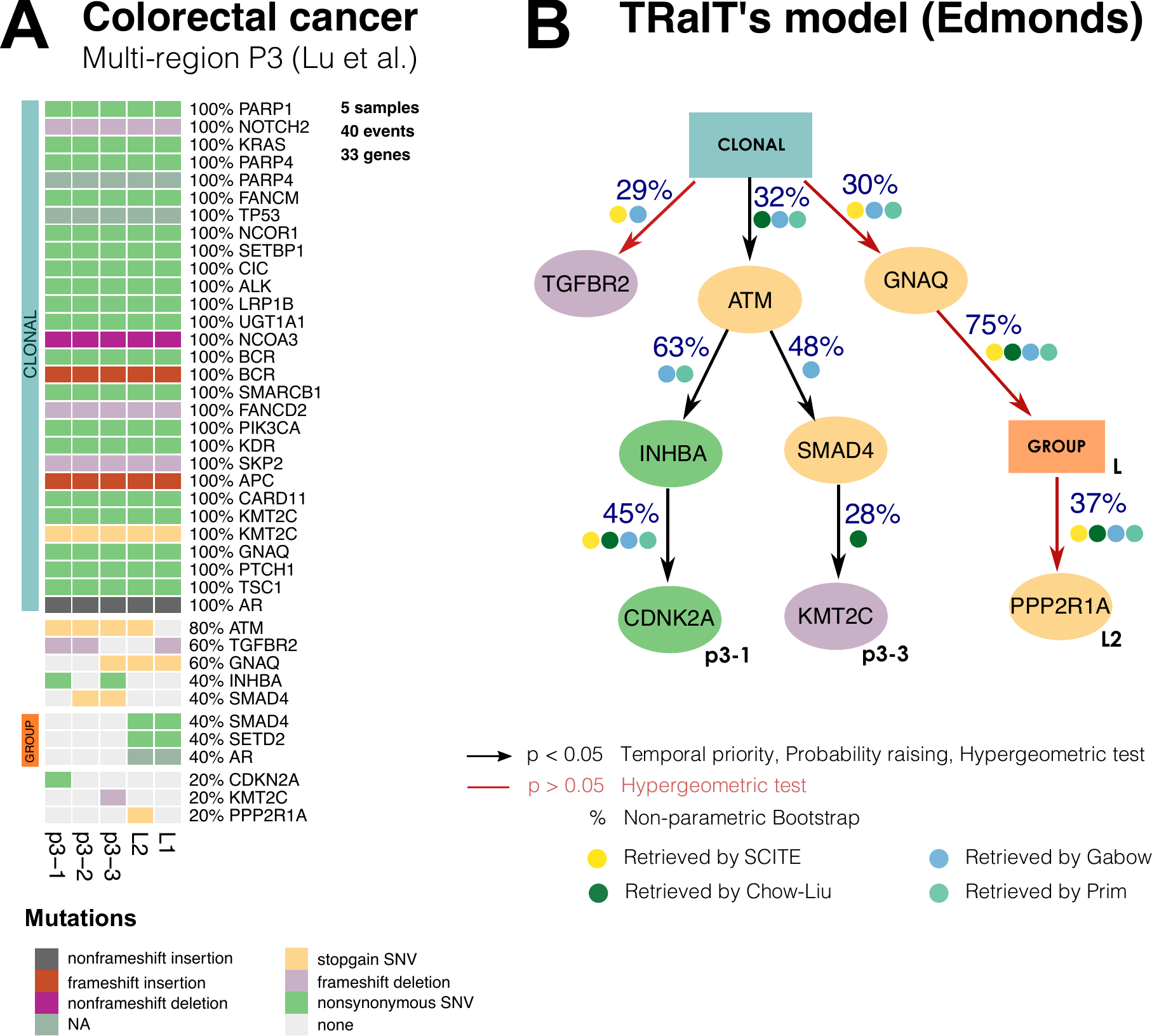}}
  \caption[Colorectal cancer multi-region data by You-Wang]
  {\textbf{A}. Multi-region sequencing data for a MSI-high colorectal
    cancer  from \cite{lu2016colorectal}, with  three regions  of the
    primary cancer: \texttt{p3-1}, \texttt{p3-2} and \texttt{p3-3}, and two of
    one metastasis: \texttt{L-1} and \texttt{L-2}. To use this data with
    \TRAIT{} we merge mutations   occur in the same samples, obtaining a {clonal} group
    of $34$ mutations and a  {sublclonal} group.
    \textbf{B}. The model obtained by \edm{} including confidence
    measures, and the overlap in the predicted ordering obtained by
    \SCITE{}, \chl{}, \gbw{} and  \prim{} (Supplementary Figure S21).
    All edges, in all models, are statistically significant for
    conditions (\ref{eq:sup}). Four of the predicted ordering
    relations are consistently found across all \TRAIT{}'s algorithm,
    which gives a high-confidence explanation for the formation of the
    \texttt{L2} metastasis. This finding is also in agreement with
    predictions by \SCITE{} (Supplementary Figure S22).}
\label{fig:mreg-crc}
\end{figure}

\textcolor{black}{We applied  \TRAIT{} to  47  {nonsynonymous point mutations} and 11  {indels} 
detected via targeted sequencing in patient \texttt{P3} of \cite{lu2016colorectal}.
This patient has been diagnosed with a moderately-differentiated MSI-high colorectal cancer, for
which  3 samples are collected from the  primary tumour (\texttt{P3-1}, \texttt{P3-2},
and \texttt{P3-3}) and two from a right hepatic lobe metastasis \texttt{L-1} and
\texttt{L-2}  (Figure \ref{fig:mreg-crc}\textsf{A}).} To prepare the data for our analyses,  we first grouped  mutations occurring in the same  regions. We  obtained: $(a)$ a \textsf{clonal} group of 
 $34$ mutations detected in all  samples $(b)$ a \textsf{subclonal} group of $3$ mutations private
 to the metastatic  regions, and $(c)$ $8$ mutations with distinct mutational profiles. The clonal
 group contains mutations in  key colorectal driver genes such as  \texttt{APC}, \texttt{KRAS},
\texttt{PIK3CA} and \texttt{TP53} \cite{Caravagna2016E4025},

\edm{}'s model predicts branching evolution and  high levels
of ITH among the subclonal populations, consistently with the
original phylogenetic analysis by Lu et al.  \cite{lu2016colorectal} (Figure
\ref{fig:mreg-crc}\textsf{B}).
In particular, the subclonal trajectory that \textcolor{black}{characterizes} the primary
regions is initiated by  a  {stopgain SNV} in the DNA damage
repair gene  \texttt{ATM}, whereas  the  {subclonal metastatic}
expansion seems to  originate by a stopgain SNV in \texttt{GNAQ}, a gene reponsible for diffusion in many tumour types 
\cite{kim2014gnaq}. The model also pictures  two distinct trajectories with  different mutations in
\texttt{SMAD4}: a nonsynonimous SNV  in group \texttt{L}, and a stopgain SNV in two regions of the primary. 
Interestingly, \texttt{SMAD4}  regulates cell proliferation, differentiation and apoptosis \cite{alazzouzi2005smad4}, and
 its loss is correlated with colorectal metastases  \cite{li2011roles}. 

\textcolor{black}{We applied \SCITE{} to the same data (Supplementary Figure S22), and
compared it to \edm{}. Both  models depict the same history for the metastatic
branch, but different  tumour initiation: \SCITE{} places the  \texttt{ATM} mutation on top of
the clonal mutations, which appear ordered in a linear chain of $34$ events. 
However, this ordering is uncertain because  \SCITE{}'s posterior is multi-modal (i.e.,
several orderings have  the same likelihood; Supplementary Figure 22).} 
Further comments on the results, and outputs from other
algorithms are available Supplementary Material (Supplementary Figure 21).

\subsection*{Analysis of patient-derived SCS data for a  triple-negative breast cancer}
\label{sect:scs-tnbreast}

\begin{figure*}[t]
\centerline{\includegraphics[width=1\textwidth]{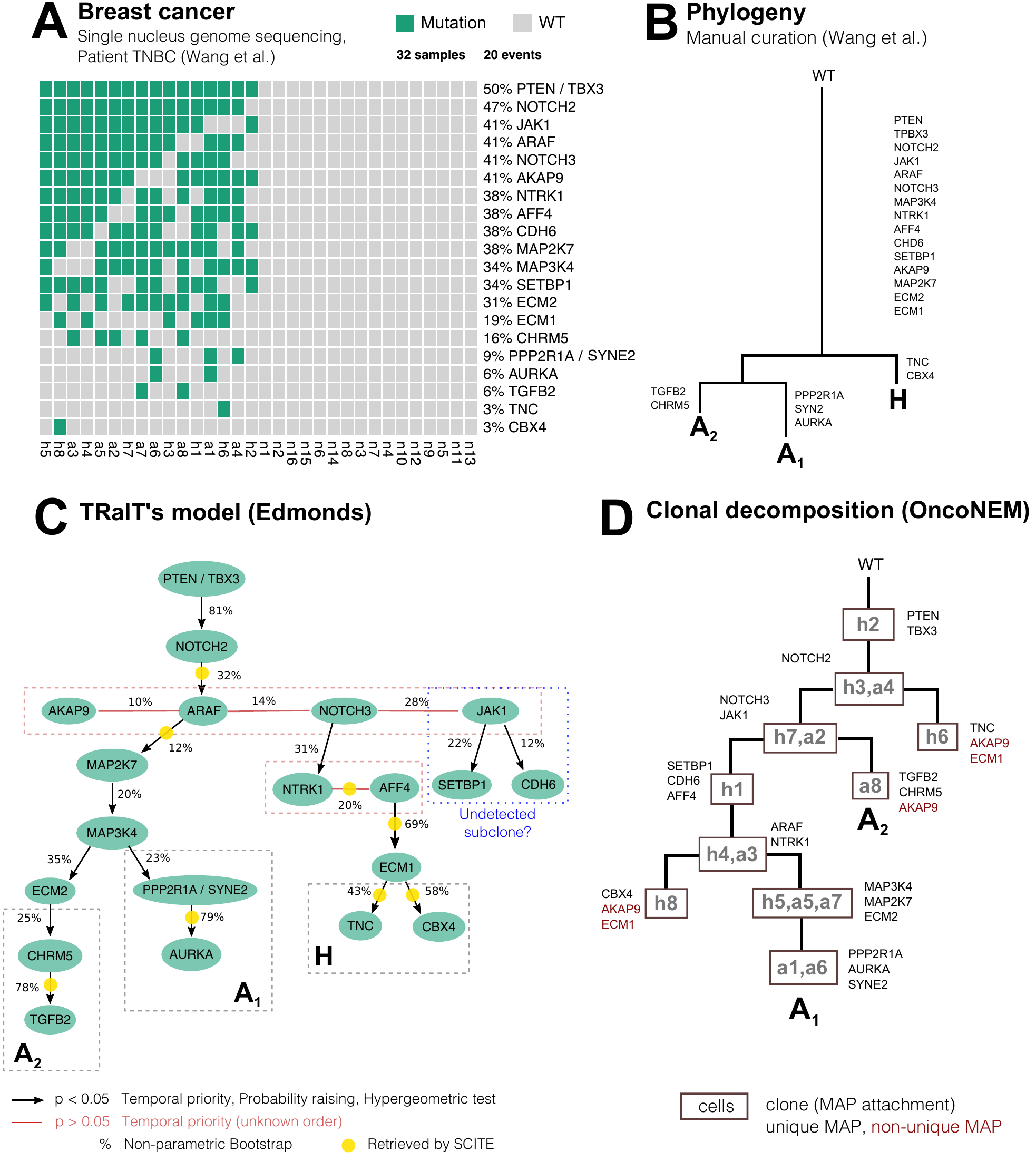}}
  \caption[Estrogen-receptor positive breast cancer (single-cell)]
  {\textbf{A.} Input data from single-nucleus sequencing  of 32 cells from a
      triple-negative  breast cancer \cite{wang2014clonal}. 
    \textcolor{black}{As the rate of missing values in the original data was around $1\%$, the authors set all missing data points equal to $0$;
    in the dataset, allelic dropout  is equal to $9.73 \times 10^{-2}$, and false
    discovery equal to $1.24 \times 10^{-6}$.}
    \textbf{B} Phylogenetic tree manually curated
      in \cite{wang2014clonal}. Mutations
    are annotated to the trunk if they are ubiquitous across cells and
    a bulk control sample.
    Subclonal mutations  appearing only in more than one
    cell.
    \textbf{C}. Mutational graph obtained with \edm{} algorithm;
    p-values are obtained by 3 tests for conditions  (\ref{eq:sup})
    and overlap  (hypergeometric test), and edges annotated with a
    posteriori non-parametric bootstrap scores ($100$ estimates). For
    these data, all \TRAIT{}'s algorithms return trees (Supplementary
    Figure S17-18), consistently with the manually curated phylogeny (A).
    Most edges are highly confident ($p < 0.05$), except for groups of
    variables with the same  frequency which have unknown ordering
    (red edges). The ordering of mutations in subclones  \texttt{A$_1$},
    \texttt{A$_2$} and tumour initiation has  high bootstrap estimates
    ($> 75\%$).
    Yellow circles mark the edges retrieved also by \SCITE{}. 
    \textbf{D}. We also performed clonal tree inference
    with \OncoNEM{}, which predicts $10$  clones. Mutations are
    assigned to clones via  \emph{maximum a posteriori} estimates. The
    mutational orderings of the early clonal expansion of the tumour
    and of most of the late subclonal events are consistent with
    \TRAIT{}'s prediction.}
\label{fig:single_cell_1}
\end{figure*}

We applied \TRAIT{} to the triple-negative breast cancer patient \texttt{TNBC} of \cite{wang2014clonal}.
The input data consists of single-nucleus exome sequencing of $32$ cells: $8$
{aneuploid} (\texttt{A}) cells, $8$  {hypodiploid}
(\texttt{H}) cells and $16$ normal cells (\texttt{N}) (Figure
\ref{fig:single_cell_1}\textsf{A}). Wang et al considered clonal all mutations detected  in a control bulk sample and in the majority of the
single cells, and as  subclonal those undetected in the bulk  \cite{wang2014clonal};  all mutations were then used to manually curate a
  phylogenetic tree (Figure \ref{fig:single_cell_1}\textsf{B}).

We run \TRAIT{} on all single cells, with nonsynonymous point
  mutations annotated in $22$ genes, and set
$\epos = 1.24 \times 10^{-6}$ and $\eneg = 9.73 \times 10^{-2}$ as
suggested in \cite{wang2014clonal}. All \TRAIT{}'s algorithms return
tree topologies (Supplementary Figures 17--18) \revfinal{\sout{indicating  lack of
violations of the ISA}}; Figure \ref{fig:single_cell_1}\textsf{C}  shows the
model obtained with \edm{}. We integrate the analysis  by applying \SCITE{} to the same data, and by computing  prevalence and evolutionary relations of putative clones with \OncoNEM{}  as well (Figure \ref{fig:single_cell_1}\textsf{D}).

\textcolor{black}{\TRAIT{} provides a  finer resolution to the original analysis
  by Wang et al \cite{wang2014clonal}, and retrieves  gradual
accumulation of point mutations thorough  tumour evolution, which highlight progressive DNA
repair and replication deregulation.}  The model also predicts  high-confidence branching evolution
patterns consistent with subclones  \texttt{A$_1$} (\texttt{PPP2R1A},
\texttt{SYNE2} and \texttt{AURKA}),  \texttt{A$_2$}
(\texttt{ECM2}, \texttt{CHRM5} and \texttt{TGFB2}), and
\texttt{H} (\texttt{NRRK1}, \texttt{AFF4}, \texttt{ECM1},
\texttt{CBX4}), and  provides an explicit ordering among clonal  mutations in 
\texttt{PTEN}, \texttt{TBX3} and \texttt{NOTCH2}, which  trigger  tumour initiation.
Interestingly,  \TRAIT{} also allows to formulate new hypotheses about a possibly 
undetected subclone  with  private mutations   in \texttt{JAK1}, \texttt{SETBP1} and
\texttt{CDH6}.  Finally,  we note that  that temporal ordering among
mutations in  \texttt{ARAF}, \texttt{AKAP9}, \texttt{NOTCH3} and \texttt{JAK1} cannot
be retrieved, since these events have  the same
marginal probability in these data.

By applying \SCITE{} to these  data with the same noise rates, \textcolor{black}{
we retrieved 10.000 equivalently optimal trees. The overlap  between the first of the returned
 trees (Supplementary Figure S19)  and ours is poor (8 out of 19 edges), and \SCITE{}'s models contain  a long linear  chain of 13 truncal mutations.
Clonal deconvolution analysis via \OncoNEM{} allowed us to   detect  10 clones, their lineages and evolutionary relations.  This analysis
is in stronger agreement with ours, and the   estimated  mutational ordering  obtained by assigning  mutations to
clones (via maximum a posteriori, as suggested in \cite{ross2016onconem})  largely overlaps with
 \TRAIT{}'s predictions. }This is particularly evident for early
events, and for most of the late subclonal ones, exception made for
subclone \texttt{H}, which is not detected by \OncoNEM{}.
These results prove that concerted application of tools for mutational
and clonal trees inference can provide a picture of ITH at an
unprecedented resolution.

\section*{Discussion}
\label{sect:discussion-pre}
 
 \revfinal{In this paper we have introduced \TRAIT{}, a computational approach for the inference of cancer evolution models in single tumours. 
 \TRAIT{}'s expressive framework allows to reconstruct models beyond standard trees, such as forests, which capture different modalities of tumour initiation (e.g., by multiple cells of  origin, or by events missing in available genomic data, such as epigenetic states) and, under certain conditions of data and parameters, confluences. 
Future works will exploit this latter feature to define a comprehensive modelling framework that accounts for explicit violations of the ISA, in order to model further evolutionary phenomena, such as convergent (parallel) evolution and back mutations  \cite{navin2016genotyping}.} 

\revfinal{
\TRAIT{} is based on a binary representation of input data, for both multi-region and single-cell sequencing data.
We  comment on this design choice concerning the case of multi-region bulk data, because most  methods that process bulk data use allelic frequencies and cancer cell fractions
to deconvolve the clonal composition of a tumor  (see, e.g., \cite{deshwar2015phylowgs,el2016inferring,jiang2016assessing}).
In this respect,  allele frequency-derived inputs provide higher-resolution estimates of the temporal orderings among samples. In fact,  if two mutations co-occur  in the same set of samples,    their relative temporal ordering cannot be determined from a binary input, while this might be possible from  their cancer cell fractions. However, despite the lower resolution, a binary representation  is still a viable option in multi-region analyses.}

\revfinal{First, binary data can describe the presence or absence of a wide range of covariates, which otherwise might be difficult or impossible to represent  with allele-frequencies or cancer cell fractions. These include, for instance, complex structural re-arrangements, structural variants, epigenetic modifications, over/under gene expression states and high-level pathway information. The integration of such heterogeneous data types and measurements will be essential to deliver an effective {multi-level representation} of the life history of individual tumours. Methods that strictly rely on allelic frequencies might need to be extended to accommodate such data types.}

 \revfinal{Second, binary inputs can be used to promptly analyse targeted sequencing panels, whereas the estimation of  subclonal clusters from allele frequencies (i.e., via subclonal deconvolution)  requires at least  high-depth whole-exome sequencing data to produce reliable results. While it is true that whole-exome and whole-genome assays are becoming increasingly common, many large-scale genomic studies are still relying on targeted sequencing (see, e.g., \cite{Turajlic:2018:_trecerx,turajlic2018tracking}), especially in the clinical setting. A prominent example are  assays for longitudinal sampling of circulating tumour DNA during therapy monitoring, which  often consist of deep-sequencing target panels derived from the composition of a primary tumour (see, e.g., \cite{abbosh2017phylogenetic}).}

\revfinal{Finally, binary inputs can be obtained for both bulk and single-cell sequencing data, and this in turn allows to use the same framework to study cancer evolution from both data types. This is innovative, and in the future integrative methods might draw inspiration from our approach.}

\section*{\textcolor{black}{Conclusions}}
\label{sect:discussion}

\textcolor{black}{Intra-tumour heterogeneity is a  product of the interplay arising from competition,
selection and neutral evolution of cancer  subpopulations, and  is one of the major causes of {drug resistance},
{therapy failure} and {relapse}  
\cite{nik2012life,gillies2012evolutionary,burrell2013causes,vogelstein2013cancer,sottoriva2013intratumor}.}
For this reason, the choice of the appropriate statistical approach to
take full advantage of the increasing resolution  of
genomic data is key to produce {predictive models} of tumour evolution
with {translational relevance}.

We have here introduced \TRAIT{}, a framework for the efficient
reconstruction of single tumour evolution from multiple-sample
sequencing data. Thanks to the simplicity of the underlying
theoretical framework, \TRAIT{} displays significant advancements in
terms of  {robustness},  {expressivity},  {data  integration} and {computational complexity}.
 \TRAIT{} can process both multi-region and SCS data (separately), and its optimal algorithms maintain a low computational burden compare to alternative tools. \TRAIT{}'s assumptions to model accumulation phenomena lead to  {accurate} and {robust estimate} of temporal orderings, also in presence of noisy data.  

\revfinal{We position TRAIT in a very precise niche in the landscape of tools for cancer evolution reconstruction, i.e., that of methods for the inference of mutational trees/graphs (not clonal or phylogenetic trees), from binary data (alteration present/absent), and supporting both multi-region bulk and single-cell sequencing data. 
We advocate the use of \TRAIT{}  as complementary to tools for clonal tree inference, in a joint effort to quantify the extent of ITH, as shown in the case study on triple negative breast cancer.} 

\revfinal{\sout{The binary data used in  \TRAIT{} for both multi-region and SCS allows one to annotated a broader set of genomic alterations compared to tools based on allelic frequencies, such as  high-level information on pathways, hallmarks, phenotypic-triggering lesions or
epigenetic states. This can be used to retrieve a {multi-level representation} of  the life history of individual tumours, and can be combined with more complex statistical methods that detect  patterns of repeated  evolution across patients.) }}

\revfinal{\sout{To conclude, we advocate the use of \TRAIT{}  as complementary to
tools for clonal tree inference, in a joint effort to quantify the
extent of ITH, as shown in the case study on triple negative breast
cancer.}}

\section*{Methods}
\label{sect:methods}

\subsection*{Input Data and Data Types}
\label{sect:input-data-types}

\TRAIT{} processes an
input binary matrix $D$ with $n$ columns and $m$ rows. $D$ stores $n$
binary variables (somatic mutations, CNAs, epigenetic states, etc.)
detected across $m$ samples (single cells or multi-region samples) (Figure \ref{fig:framework}A).
One can annotate data at different resolutions: for instance, one can
distinguish mutations by type (missense vs truncating), position, or
context (\texttt{G>T} vs \texttt{G>A}), or can just annotate a general
``mutation'' status. The same applies
for copy numbers, which can be annotated at the focal, cytoband or
arm-level. In general, if an entry in $D$ is 1, then the associated
variable is detected in the sample. 

In our framework we cannot disentangle the temporal ordering between events that occur
in the same set of samples. These  will be grouped by  \TRAIT{}  in a new ``aggregate'' node,
prior to the inference (Figure \ref{fig:framework}B). \textcolor{black}{\TRAIT{} does not explicitly account for {back
    mutations} due to loss of heterozygosity. Yet, the information on these events can be used to
    prepare input data if one matches the copy number state to the presence of mutations. 
    By merging these events we can  retrieve their temporal position
  in the output graph (Supplementary Figure S23).}
  
\TRAIT{} supports both multi-region and SCS data. As we expect $D$ to
contain {noisy observations} of the {unknown true genotypes}, 
the algorithms can be informed of {false positives and negatives}  rates ($\epos \geq0$ and
$\eneg \geq0$). \TRAIT{} does not implement noise learning strategies, similarly to \OncoNEM{}
 \cite{jahn2016tree}. \textcolor{black}{This choice is sensitive  if the algorithms show stable
performance for slight variations in the input noise rates,  especially when
reasonable estimates of $\epos$  and $\eneg$ can be known a priori.} This feature  allows
\TRAIT{} to be computationally more efficient, as it avoids to include a noise learning routine in the fit.
{Missing data}, instead, are handled by  a standard
{Expectation Maximisation} approach to impute missing values: for every complete dataset obtained, 
the fit is repeated  and the  model that maximises the likelihood across  all runs is returned. 

\subsection*{\TRAIT{}'s Procedure} 

\textcolor{black}{All \TRAIT{}'s algorithms can be summarised with a three-steps skeleton, where  the first
two steps are the same across all  algorithms. Each algorithm will return  a unique output model, 
whose   post hoc confidence can be assessed via  
cross-validation and bootstrap \cite{Caravagna2016E4025}.}

\subsubsection*{Step 1: assessment of statistical association -- Figure \ref{fig:framework}C}

We estimate  the statistical association between events by assessing two conditions inspired to Suppes' theory of
{probabilistic causation}, which is  particularly sound in modelling {cumulative} phenomena  \cite{suppes1970probabilistic}.
  
Let $p(\cdot)$ be an empirical  probability (marginal, joint,
conditional etc.) estimated from  dataset $D$.  Conditions on $(i)$
{temporal direction} and $(ii)$ {association's strength} are
assessed as follows: for every pair of variables $x$ and $y$ in $D$,
$x$ is a plausible temporally antecedent  event of $y$ if
\begin{eqnarray}
  \label{eq:sup}
  p(x) > p(y)  & \wedge &  p(y \mid x) >p (y \mid \neg x) \, .
\end{eqnarray}
The former condition acts as the {Infinite Sites Assumption}
(ISA), as we assume that {alterations are inherited} across
 cell divisions (i.e., somatic): thus, the comparison of 
marginal frequencies is a proxy to compute the relative ordering among
events. The latter condition, instead, implies statistical dependence: $p(x,y) > p(x)p(y)$ 
\cite{loohuis2014:_caprese_infer}. 

Both conditions are assessed among all variables pairs via
{non-parametric bootstrap} and a one-tailed
Mann-Whitney test:  only if both  conditions are statistically significant at some
$\alpha$-level (e.g., $0.05$), the edge connecting the variable pair
will be included in a prima-facie {direct graph} $G_\mathsf{pf}$.
Edges in $G_\mathsf{pf}$ are candidate to be selected in the final output model, and thus
we are reducing  the search space via  the  the above conditions, which are  necessary but not
sufficient. \textcolor{black}{These conditions
  have been previously used to define causal approaches for cancer
  progression
  \cite{ramazzotti2015capri,Caravagna2016E4025}; see further discussion in Supplementary Material}.  
This step has asymptotic complexity ${\cal O}((nm)^2 \times B)$ where $B$
is the  cost of  bootstrap and {hypothesis testing} on each entry in $D$. Notice that 
this procedure can  create disconnected components.

\subsubsection*{Step 2: loop removal -- Figure \ref{fig:framework}D}

\GPF{} can contain loops, which we have to remove to model an accumulation process.
Loops  may arise  when an arc between a pair of nodes cannot be unequivocally directed, e.g., 
due to small sample size which leads to uncertain bootstrap estimations. 
\TRAIT{} renders acyclic  \GPF{}  by using heuristic strategies
that remove less confident edges (see \cite{ramazzotti2015capri}); the output 
produced is a new graph \GI{}. 

\subsubsection*{Step 3: reconstruction of the output model -- Figure \ref{fig:framework}E--F}

We render   \GI{}  a  weighted graph by annotating its edges  via
information-theoretic measures such as {point-wise mutual
  information} and the like. Then,  we can
exploit  4 different off-the-shelf algorithms to reconstruct an output
model \GMO{} from  \GI{}.   \GMO{} will be either a tree or a forest with
multiple roots, and the complexity of this step depends on
the adopted algorithm. Notably, all  algorithms currently incorporated
in \TRAIT{}   have theoretically-optimal worst-case polynomial
complexity. We  describe two of them (\edm{}  and \chl{}),  and leave the description of the
other techniques (\gbw{} and  \prim{}) to the Supplementary Material.

\begin{itemize}
\item \edm{}    is an  algorithm for the inference of {weighted directed minimum
    spanning trees} \cite{edmonds1968optimum}: it scan  \GI{} to identify the
  tree that maximises the edges' weights. Spanning trees have been previously applied to 
    cancer \cite{ElKebir:2015:_reconstruction,Popic:2015:_fast}.  Yet, \TRAIT{} is the only framework to constraint  
    spanning trees    by condition (\ref{eq:sup});
    
\item  \chl{}'s algorithm  is a method to compute a    factorisation of a joint
  distribution over the input variables \cite{chow1968approximating}.   \chl{} reconstructs 
   {undirected} trees by definition; we assign the direction to each 
   edge so that the event with higher marginal probability is
   on top, mirroring condition (\ref{eq:sup}). 
   Confluences in $G_\mathsf{MO}$ can emerge \revfinal{\sout{by this construction, modelling possible violations of the ISA}} under certain conditions of the observed probabilities, which \revfinal{account for the uncertainty on the temporal precedence among events (\revfinal{technically, in such cases we reconstruct direct acyclic graphs, DAGs -- }see the Supplementary Material for details)}.
  \end{itemize}
In all \TRAIT{}'s  algorithms, if \GI{}
 includes $k$ disconnected components, then the output model $G_\mathsf{MO}$
  will include $k$ disconnected trees \revfinal{\sout{and/or DAGs}}.

\textcolor{black}{In term of complexity, we note that all \TRAIT{}'s algorithms are optimal
  polynomial-time algorithmic solutions to each of their corresponding
  combinatorial problems.  Thus, they {scale well with sample
  size}, a problem sometimes observed with Bayesian approaches that
cannot compute a full posterior on the model parameters. }
\textcolor{black}{Quantitative assessment of  \TRAIT{}'s scalability with large datasets is provided as Supplementary Material (Supplementary Table 7), where we show that \rev{many thousands of cells} can be processed in a few seconds}. 

\subsection*{Tumour evolution scenarios}
\label{sect:working-scenarios}

\TRAIT{} can   infer mutational graphs in \rev{the following scenarios (see Figure \ref{fig:approach}D):}
\begin{enumerate}
\item {Branching evolution} (including  {linear evolution} as subcase): in this case \TRAIT{} will return
  a tree  with one root and zero disconnected components. 

\item Presence of {confounding factors} in $D$ (e.g., miscalled
  mutations): \TRAIT{} will reconstruct a model with disconnected individual
  nodes.

\item {Polyclonal origin} due   to {multiple cells of tumour origin}, or to upstream events
  triggering tumour development that missing in $D$ \textcolor{black}{(e.g., epigenetic events)}:
  \TRAIT{} will return models with disconnected components (i.e., forests).

\revfinal{\sout{ (iv) Generalised {convergent evolution}. This can happen also because of the data that we have preprocessed.
For instance, when \TRAIT{} mutations are not distinguished by nucleotide change and regarded as 
general ``mutations'', while the underlying true evolution contains distinct types of mutations in independent subclonal lineages.
When that is the case, the processed data might be fit with a general acyclic graph with confluences, which can be interpreted
as a violation of the ISA. We note also that only \chl{} can explicitly account  for this scenario because it first renders an
undirected graph.}}
 
\end{enumerate}

In general, we recommend to apply all \TRAIT{}'s algorithms and to
compare the output models; the creation of  a {consensus model} is an option to rank 
the edges detected across several methods, as we show in the case studies.

\section*{\textcolor{black}{Abbreviations}}
\textcolor{black}{
CNA: Copy-Number Alteration\\
CT: Clonal Tree\\ 
ISA: Infinite Sites Assumption \\
ITH: Intra-Tumour Heterogeneity \\
MSI: Micro-Satellite Instable \\
SCS: Single-Cell Sequencing \\
SNV: Single-Nucleotide Variant}

\subsection*{Availability of data and materials}
\TRAIT{} is included in the \TRONCO{} R suite for TRanslational
ONCOlogy, available at its webpage {\sf \href{http://troncopackage.org}{http://troncopackage.org}}
and mirrored at {\sf Bioconductor}. All
data used in this paper are available from the supplementary material
of \cite{wang2014clonal} and \cite{lu2016colorectal}. We provide the
source code and the input data to reproduce the case studies at: 
{\sf \href{https://github.com/BIMIB-DISCo/TRaIT}{BIMIB-DISCo/TRaIT}}. 

\subsection*{Competing interests}
The authors declare that they have no competing interests.

\subsection*{\textcolor{black}{Funding}}
\textcolor{black}{This work was partially supported by the Elixir
Italian Chapter and the SysBioNet project, a Ministero
dell'Istruzione, dell'Universit\'a e della Ricerca initiative for the
Italian Roadmap of European Strategy Forum on Research
Infrastructures.}
\\
The funding body did not play any role in the design of the study and collection, analysis, and interpretation of data and in writing the manuscript.

\subsection*{Author's contributions}
\label{sect:author-contr}

DR, AG and GC designed the algorithmic framework. DR, LDS and GC
implemented the tool. LDS carried out the simulations on synthetic
data. Data gathering was performed by DR, AG, LDS and GC. All the
authors analyzed the results and interpreted the models.  DR, AG, MA
and GC wrote the draft of the paper, which all authors reviewed and
revised.

\subsection*{Acknowledgements}
\label{sect:acknowledgements}

We thank Bud Mishra for  valuable insights on the effects of
parallel evolution to our framework. We also thank Guido Sanguinetti
and Yuanhua Huang for useful discussions on the preliminary version of
this manuscript. 

\clearpage

\bibliographystyle{unsrt}
\bibliography{bibliography}

\clearpage

\includepdf[pages=-]{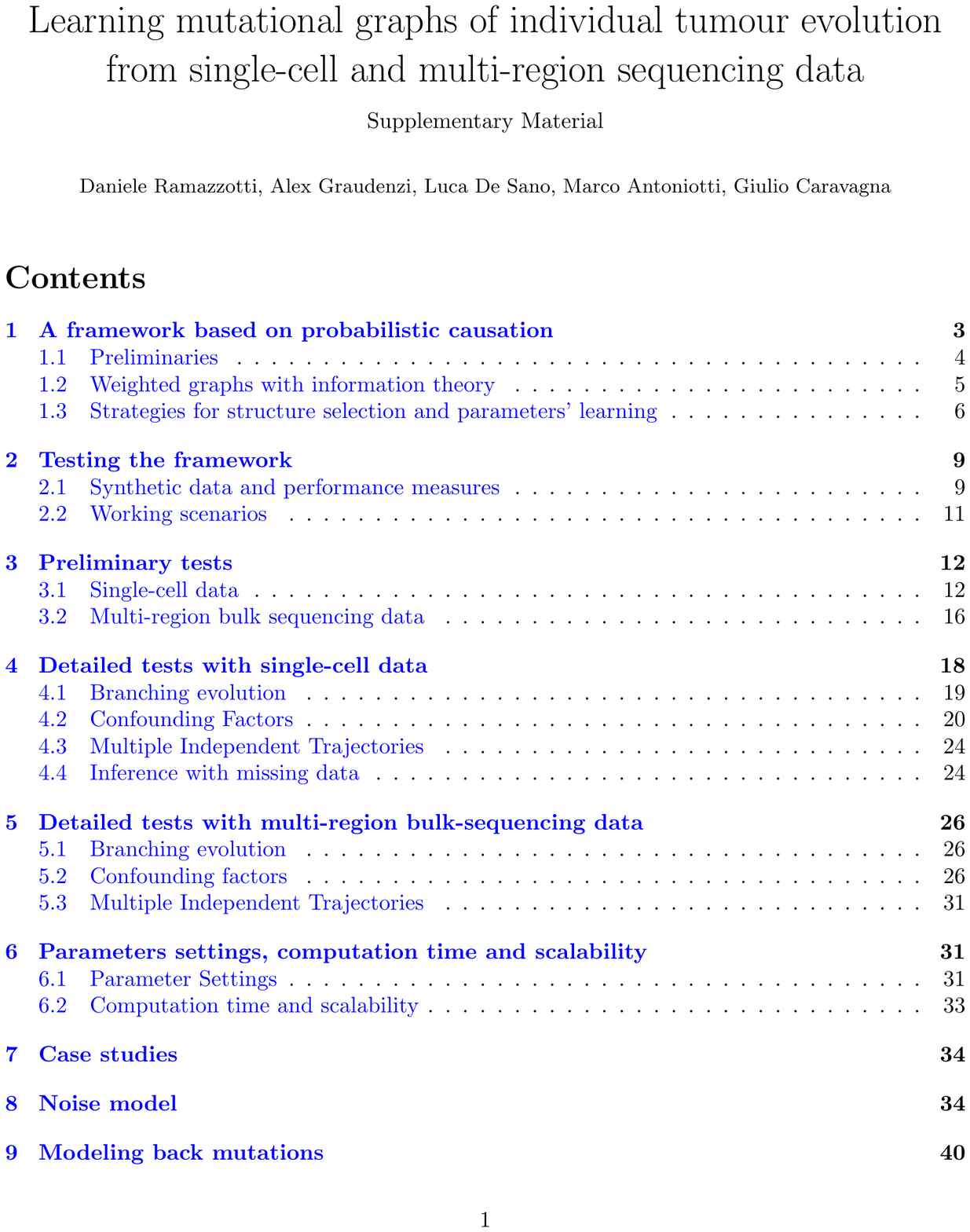}

\end{document}